\documentclass[epj]{webofc}
\usepackage[varg]{txfonts}
\usepackage{amsmath}
\usepackage{amsfonts}
\usepackage{amssymb}
\usepackage{graphicx}
\usepackage[toc,page]{appendix}

\wocname{EPJ Web of Conferences}
\woctitle{ICNFP 2017}

\begin{document}

\title{Excited vector mesons: phenomenology and predictions for a yet
unknown vector $s\bar{s}$ state with a mass of about 1.93~ GeV}
\author{M. Piotrowska\inst{1}\fnsep\thanks{\email{milena.soltysiak@op.pl}} \and
        F. Giacosa\inst{1,2}
}

\selectlanguage{english}

%
%

\institute{Institute of Physics, Jan Kochanowski University,\\\textit{ul. Swietokrzyska 15, 25-406, Kielce, Poland. }\\ 
\and
Institute for Theoretical Physics, J. W. Goethe University,\\\textit{ Max-von-Laue-Str. 1, 60438 Frankfurt, Germany.}
}

\abstract{  The firm understanding of standard quark-antiquark states (including excited states) is necessary to search for non-conventional mesons with the same quantum numbers. In this work, we study the phenomenology of two nonets of excited vector mesons, which predominantly correspond to radially excited vector mesons with quantum numbers $n$ $^{2S+1}L_{J}=2$ $^{3}S_{1}$ and to orbitally excited vector mesons with quantum numbers $n$~ $^{2S+1}L_{J}=1$ $^{3}D_{1}$. We evaluate the decays of these mesons into two pseudoscalar mesons and into a pseudoscalar and a ground-state vector meson by making use of a relativistic quantum field theoretical model based on flavor symmetry. Moreover, we also study the radiative decays into a photon and a pseudoscalar meson by using vector meson dominance. We compare our results to the PDG and comment on open issues concerning the corresponding measured resonances. Within our approach, we are also able to make predictions for a not-yet discovered $s\bar{s}$ state in the $n$ $^{2S+1}L_{J}=1$ $^{3}D_{1}$ nonet, which has a mass of about 1.93 GeV. This resonance can be searched in the upcoming GlueX and CLAS12 experiments which take place at the Jefferson Lab.
}
\maketitle

\section{Introduction}

In the last decades many theoretical calculations have appeared and numerous
experimental observations have been made to explain the structure of matter.
This research led to the postulate and verification of quarks, which
interact through gluons and build up \textquotedblleft
white\textquotedblright\ hadrons. These strong interactions are described by
Quantum Chromodynamics (QCD). We consider here bosonic hadrons, the so$-$%
called mesons; a significant number of mesons consists of a quark-antiquark $%
(q\bar{q})$ pair. These are the \textquotedblleft conventional
mesons\textquotedblright , for more details see the review on the quark
model in Ref. \cite{godfrey}. Besides them, there is also a group of
non-conventional mesons such as four$-$quark or gluonic states. Nowadays new
evidence of their existence is emerging, see Refs. \cite{amsler, klempt}. 

Ground-state vector mesons with the principal quantum number $n=1$, spacial
angular momentum $L=0$ and spin $S=1$ are very well known. On the contrary,
the understanding of the nature of excited states is not yet so advanced.
However, a good description of the excited states is necessary to prove the
validity of the conventional quark$-$antiquark assignment and to understand
the nature of non$-$conventional states beyond this standard picture.\newline
In this work we study the phenomenology of two types of excited, quark$-$%
antiquark vector mesons (build up from the light $u$, $d$ and $s$ quarks) in
the low$-$energy regime, for which there are numerous experimental
observations. The first group is identified with the states $\{\rho
(1450),K^{\ast }(1410),\omega (1420),\phi (1680)\}$ characterized by the
principal number $n=2$, spacial angular momentum $L=0$ and spin $S=1$ (in
the old spectroscopy notation $n$ $^{2S+1}L_{J}=2$ $^{3}S_{1}$). They are
(predominantly) radially excited vector mesons. The second group is
identified with the states $\{\rho (1700),K^{\ast }(1680),\omega (1650),\phi
(???)\}$ characterized by the principal number $n=1$, spacial angular
momentum $L=2$ and spin $S=1$ (which corresponds to $n$~ $^{2S+1}L_{J}=1$ $%
^{3}D_{1}$). They are (predominantly) orbitally excited vector mesons. The
last missing state in this nonet, $\phi (???)$, needs still experimental
confirmation. By using our approach we make predictions for this putative
state.\newline
This paper, based on the publication \cite{mpio}, is organized as it
follows: in Sec. \ref{Sec.2} we describe the Quantum Field Theoretical (QFT)
relativistic Lagrangian of our model; in Sec. \ref{Sec.3} we present the
results for two types of decays, strong and radiative ones, and we show the
predictions for a putative $s\bar{s}$ state in the nonet of orbitally
excited vector mesons. Then, in Sec. \ref{Sec.4} we present our conclusions.

\section{The theoretical model}

\label{Sec.2} In order to study the decays of excited vector mesons we
utilize an effective QFT relativistic model. In the past, such models have
been successfully applied to tensor mesons \cite{tensor}, pseudotensor
mesons \cite{pseudotensor}, as well as to pseudovector \cite{pseudovector}
and (nontrivial) scalar mesons \cite{gutsche, gutsche2}. The Lagrangian of
our model fulfills the following conditions: it is invariant under flavour
transformation $U(3)_{V}$, parity $P$, and charge conjugation $C$ and it
contains four mesonic nonets in form of $3\times 3$ matrices with the
elements corresponding to $q\bar{q}$ states, see below. Its explicit form
reads: 
\begin{equation}
\mathcal{L}=ig_{EPP}Tr\left( [\partial ^{\mu }P,V_{E,\mu }]P\right)
+ig_{DPP}Tr\left( [\partial ^{\mu }P,V_{D,\mu }]P\right) +g_{EVP}Tr\left( 
\tilde{V}_{E}^{\mu \nu }\{V_{\mu \nu },P\}\right) +g_{DVP}Tr\left( \tilde{V}%
_{D}^{\mu \nu }\{V_{\mu \nu },P\}\right) \text{ .}  \label{Lag}
\end{equation}%
The first two terms describe decays of the type $V_{R}\rightarrow PP$ and
the last to terms decays of the type $V_{R}\rightarrow VP$ with $R=E,D$.
Moreover, $[.,.]$ and $\{.,.\}$ are the usual commutator and anticommutator,
respectively, $V^{\mu \nu }=\partial ^{\mu }V^{\nu }-\partial ^{\nu }V^{\mu }
$ and $\tilde{V}_{R}^{\mu \nu }=\frac{1}{2}\epsilon ^{\mu \nu \alpha \beta
}(\partial _{\alpha }V_{R,\beta }-\partial _{\beta }V_{R,\alpha })$ (the
latter is the the dual field defined in the standard way). In Eq. (\ref{Lag}%
), $P$ stands for the matrix of pseudoscalar mesons and it is identified
with the states $\{\pi ,K,\eta (547),\eta ^{\prime }(958)\}$, $V_{\mu }$
describes the matrix of ground$-$state vector mesons with associated
resonances $\{\rho (770),K^{\ast }(892),\omega {782,}\phi (1020)\}$, $%
V_{E,\mu }$ stands for the nonet of radially excited vector mesons $\{\rho
(1450),K^{\ast }(1410),\omega (1420),\phi (1680)\}$ and finally, $V_{D,\mu }$
stands for the nonet of orbitally excited vector mesons $\{\rho
(1700),K^{\ast }(1680),\omega (1650),\phi (???)\}$. Moreover, in our
relativistic Lagrangian we have four coupling constants, $%
g_{EPP},g_{DPP},g_{EVP},g_{DVP}$, which have been determined in Ref. \cite%
{mpio} by using well-known experimental data listed in the Particle Data
Group (PDG) \cite{pdg}. [Note, in order to determine the errors for the
coupling constants, the uncertainties on the masses of resonances were not
taken into account. For details on how to determine the coupling constants
together with the errors, see Ref. \cite{mpio}]. The numerical values of the
coupling constants are reported in Tab 1. 
\begin{table}[h]
\caption{Results for coupling constants determined by our model.}\centering
\par
\renewcommand{\arraystretch}{1.25} 
\begin{tabular}{c|c}
coupling constant & value \\ \hline
$g_{EPP}$ & $3.66 \pm 0.4$ \\ 
$g_{EVP}$ & $18.4 \pm 3.8$ \\ 
$g_{DPP}$ & $7.15 \pm 0.94$ \\ 
$g_{DVP}$ & $16.5 \pm 3.5$ \\ \hline
\end{tabular}%
\end{table}
\newpage By using a standard QFT calculation, we obtain simple formulas for
two strong decay channels $(V_{R}\rightarrow PP,V_{R}\rightarrow VP)$ and
one radiative decay channel $(V_{R}\rightarrow \gamma P)$. The tree-level
decay widths of a resonance $R=\{E,D\}$ read: 
\begin{equation}
\Gamma _{R\rightarrow PP}=s_{RPP}\frac{|\vec{k}|^{3}}{6\pi m_{R}^{2}}\left( 
\frac{g_{RPP}}{2}\lambda _{RPP}\right) ^{2}\text{ ,}\hspace{0.5cm}\Gamma
_{R\rightarrow VP}=s_{RVP}\frac{|\vec{k}|^{3}}{12\pi }\left( \frac{g_{RVP}}{2%
}\lambda _{RVP}\right) ^{2}\text{ ,}  \label{sdec}
\end{equation}%
\begin{equation}
\Gamma _{R\rightarrow \gamma P}=\frac{|\vec{k}|^{3}}{12\pi }\left( \frac{%
g_{RVP}}{2}\frac{e_{0}}{g_{\rho }}\lambda _{R\gamma P}\right) ^{2}\text{ ,}
\label{rgammap}
\end{equation}%
where $\vec{k}$ is the value of the three$-$momentum of one emitted particle
from the decaying resonance with mass $m_{R}$, while the masses of the
emitted particles are $m_{a}$ and $m_{b}$, respectively. The coefficients $%
s_{RPP}$ and $s_{RVP}$ are symmetry factors determined by isospin. Moreover, 
$\lambda _{RPP}$, $\lambda _{RVP}$ and $\lambda _{R\gamma P}$ are
coefficients arising from expanding the traces in our Lagrangian (see tables
2-4).

\section{Results}

\label{Sec.3} This section presents the results for the decays of nonets of
radially and orbitally excited vector mesons. In Sec. \ref{sec3.1} we
concentrate on strong decays of these resonances, while in Sec. \ref{sec3.2}
we show the results for radial decays. In both cases the results are
presented in summarizing tables. In total, we determined the decay widths
for $64$ different decay channels. Then, in Sec. \ref{sec3.3} we focus on
the last missing state of the $1^{3}D_{1}$ nonet, which we called $\phi
(1930)$.

\subsection{Strong decays of excited vector mesons}

\label{sec3.1}

We calculate the strong decays of excited vector resonances into two
pseudoscalar mesons (PP) and into a ground-state vector mesons and a
pseudoscalar mesons (VP). In Tab. 2 we present the results for the nonet of
radially excited vector mesons, $\{\rho (1450),K^{\ast }(1410),\omega
(1420,\phi (1680))\}$, and in Tab. 3 for orbitally excited vector mesons, $%
\{\rho (1700),K^{\ast }(1680),\omega (1650),\phi (???)=\phi (1930)\}$.%
\newline
The tables are organized as follows. For each single decay channel we assign
the corresponding symmetry factor $s$ and $\lambda $ which we obtain from
the Lagrangian of the model (\ref{Lag}). Then, by using Eq. (\ref{sdec}) we
calculate the decay widths. As a next step we compared our results with data
listed in the PDG \cite{pdg}. Unfortunately, there is no experimental
results for every decay channel. In both tables 2 and 3 we use $``+"$ for a
decay process which has been experimentally observed and $``-"$ for those
which has not been seen in experiments. The symbol $``\ast "$ means that the
resonance has not yet been discovered.

\begin{table}[h]
\caption{Symmetry factors, amplitude's coefficients, and decay widths of
strong decays of (predominantly) radially excited vector mesons.}\centering
\par
\renewcommand{\arraystretch}{1.25} 
\begin{tabular}{c|c|c|c|c}
\hline
\multicolumn{5}{c}{\textbf{Radially excited vector mesons}} \\ \hline
Decay channel & Symmetry & Amplitude & \multicolumn{2}{c}{Decay width [MeV]}
\\ \cline{4-5}
& factor (s) & ($\lambda$) & Theory & Exp. \\ \hline
\multicolumn{5}{c}{\textbf{$V_E \rightarrow PP$}} \\ \hline
$\rho(1450) \rightarrow \bar{K}K$ & 2 & $\frac{1}{2}$ & $6.6 \pm 1.4$ & $%
<6.7 \pm 1.0$ \\ 
$\rho(1450) \rightarrow \pi \pi$ & 1 & 1 & $30.8 \pm 6.7$ & $\sim 27\pm 4$
\\ 
$K^*(1410) \rightarrow K \pi$ & 3 & $\frac{1}{2}$ & $15.3 \pm 3.3$ & $15.3
\pm 3.3$ \\ 
$K^*(1410) \rightarrow K \eta$ & 1 & $\frac{1}{2}(\cos\theta_p-\sqrt{2}\sin
\theta_p)$ & $6.9 \pm 1.5$ & - \\ 
$K^*(1410) \rightarrow K \eta^{\prime }$ & 1 & $\frac{1}{2}(\sqrt{2}\cos
\theta_p+\sin \theta_p)$ & $\approx 0$ & - \\ 
$\omega(1420) \rightarrow \bar{K}K$ & 2 & $\frac{1}{2}$ & $5.9 \pm 1.3$ & -
\\ 
$\phi(1680) \rightarrow \bar{K}K$ & 2 & $\frac{1}{\sqrt{2}}$ & $19.8 \pm 4.3$
& + \\ \hline
\multicolumn{5}{c}{\textbf{$V_E \rightarrow VP$}} \\ \hline
$\rho(1450) \rightarrow \omega \pi$ & 1 & $\frac{1}{2}$ & $74.7 \pm 31.0$ & $%
\sim 84 \pm 13$ \\ 
$\rho(1450) \rightarrow K^*(892)K$ & 4 & $\frac{1}{4}$ & $6.7 \pm 2.8$ & +
\\ 
$\rho(1450) \rightarrow \rho(770) \eta$ & 1 & $\frac{1}{2} \cos \theta_p$ & $%
9.3 \pm 3.9$ & $<16.0\pm 2.4$ \\ 
$\rho(1450) \rightarrow \rho(770) \eta^{\prime }$ & 1 & $\frac{1}{2} \sin
\theta_p$ & $\approx 0$ & - \\ 
$K^*(1410) \rightarrow K \rho$ & 3 & $\frac{1}{4}$ & $12.0 \pm 5.0$ & $<16.2
\pm 1.5$ \\ 
$K^*(1410) \rightarrow K \phi$ & 1 & $\frac{1}{2\sqrt{2}}$ & $\approx 0$ & -
\\ 
$K^*(1410) \rightarrow K \omega$ & 1 & $\frac{1}{4}$ & $3.7 \pm 1.5$ & - \\ 
$K^*(1410) \rightarrow K^*(892) \pi$ & 3 & $\frac{1}{4}$ & $28.8 \pm 12.0$ & 
$>93 \pm 8$ \\ 
$K^*(1410) \rightarrow K^*(892) \eta$ & 1 & $\frac{1}{4}(\cos \theta_p+\sqrt{%
2}\sin \theta_p)$ & $\approx 0$ & - \\ 
$K^*(1410) \rightarrow K^*(892)\eta^{\prime }$ & 2 & $\frac{1}{4}(\sqrt{2}%
\cos \theta_p-\sin \theta_p)$ & $\approx 0$ & - \\ 
$\omega(1420) \rightarrow \rho \pi$ & 3 & $\frac{1}{2}$ & $196 \pm 81$ & 
dominant \\ 
$\omega(1420) \rightarrow K^*(892)K$ & 4 & $\frac{1}{4}$ & $2.3 \pm 1.0$ & -
\\ 
$\omega(1420) \rightarrow \omega (782)\eta$ & 1 & $\frac{1}{2} \cos \theta_p$
& $4.9 \pm 2.0$ & - \\ 
$\omega(1420) \rightarrow \omega(782)\eta^{\prime }$ & 1 & $\frac{1}{2} \sin
\theta_p$ & $\approx 0$ & - \\ 
$\phi(1680) \rightarrow K \bar{K}^*$ & 4 & $\frac{1}{2\sqrt{2}}$ & $110 \pm
46$ & dominant \\ 
$\phi(1680) \rightarrow \phi (1020)\eta$ & 1 & $\frac{1}{\sqrt{2}}\sin
\theta_p$ & $12.2 \pm5.1$ & + \\ 
$\phi(1680) \rightarrow \phi (1020) \eta^{\prime }$ & 1 & $\frac{1}{\sqrt{2}}
\cos \theta_p$ & $\approx 0$ & - \\ \hline
\end{tabular}%
\end{table}

\begin{table}[h]
\caption{Symmetry factors, amplitude's coefficients, and decay widths of
strong decays of (predominantly) orbitally excited vector mesons.}%
\centering
\par
\renewcommand{\arraystretch}{1.25} 
\begin{tabular}{c|c|c|c|c}
\hline
\multicolumn{5}{c}{\textbf{Orbitally excited vector mesons}} \\ \hline
Decay channel & Symmetry & Amplitude & \multicolumn{2}{c}{Decay width [MeV]}
\\ \cline{4-5}
& factor (s) & ($\lambda$) & Theory & Exp. \\ \hline
\multicolumn{5}{c}{\textbf{$V_D \rightarrow PP$}} \\ \hline
$\rho(1700) \rightarrow \bar{K}K$ & 2 & $\frac{1}{2}$ & $40 \pm 11$ & $%
8.3^{+10}_{-8.3}$ \\ 
$\rho(1700) \rightarrow \pi \pi$ & 1 & 1 & $140 \pm 37$ & $75\pm 30$ \\ 
$K^*(1680) \rightarrow K \pi$ & 3 & $\frac{1}{2}$ & $82 \pm 22$ & $125 \pm 43
$ \\ 
$K^*(1680) \rightarrow K \eta$ & 1 & $\frac{1}{2}(\cos\theta_p-\sqrt{2}\sin
\theta_p)$ & $52 \pm 14$ & - \\ 
$K^*(1680) \rightarrow K \eta^{\prime }$ & 1 & $\frac{1}{2}(\sqrt{2}\cos
\theta_p+\sin \theta_p)$ & $0.72 \pm 0.02$ & - \\ 
$\omega(1650) \rightarrow \bar{K}K$ & 2 & $\frac{1}{2}$ & $37 \pm 10$ & - \\ 
$\phi(1930) \rightarrow \bar{K}K$ & 2 & $\frac{1}{\sqrt{2}}$ & $104 \pm 28$
& * \\ \hline
\multicolumn{5}{c}{\textbf{$V_D \rightarrow VP$}} \\ \hline
$\rho(1700) \rightarrow \omega \pi$ & 1 & $\frac{1}{2}$ & $140 \pm 59$ & $+$
\\ 
$\rho(1700) \rightarrow K^*(892)K$ & 4 & $\frac{1}{4}$ & $56 \pm 23$ & $83
\pm 66$ \\ 
$\rho(1700) \rightarrow \rho(770) \eta$ & 1 & $\frac{1}{2} \cos \theta_p$ & $%
41 \pm 17$ & $68\pm 42$ \\ 
$\rho(1700) \rightarrow \rho(770) \eta^{\prime }$ & 1 & $\frac{1}{2} \sin
\theta_p$ & $\approx 0$ & - \\ 
$K^*(1680) \rightarrow K \rho$ & 3 & $\frac{1}{4}$ & $64 \pm 27$ & $101\pm 35
$ \\ 
$K^*(1680) \rightarrow K \phi$ & 1 & $\frac{1}{2\sqrt{2}}$ & $13 \pm 6$ & -
\\ 
$K^*(1680) \rightarrow K \omega$ & 1 & $\frac{1}{4}$ & $21 \pm 9$ & - \\ 
$K^*(1680) \rightarrow K^*(892) \pi$ & 3 & $\frac{1}{4}$ & $81 \pm 34$ & $96
\pm 33$ \\ 
$K^*(1680) \rightarrow K^*(892) \eta$ & 1 & $\frac{1}{4}(\cos \theta_p+\sqrt{%
2}\sin \theta_p)$ & $0.5 \pm 0.2$ & - \\ 
$K^*(1680) \rightarrow K^*(892)\eta^{\prime }$ & 2 & $\frac{1}{4}(\sqrt{2}%
\cos \theta_p-\sin \theta_p)$ & $\approx 0$ & - \\ 
$\omega(1650) \rightarrow \rho \pi$ & 3 & $\frac{1}{2}$ & $370 \pm 156$ & $%
205 \pm 23$ \\ 
$\omega(1650) \rightarrow K^*(892)K$ & 4 & $\frac{1}{4}$ & $42 \pm 18$ & -
\\ 
$\omega(1650) \rightarrow \omega (782)\eta$ & 1 & $\frac{1}{2} \cos \theta_p$
& $32 \pm 13$ & $56\pm 30$ \\ 
$\omega(1650) \rightarrow \omega(782)\eta^{\prime }$ & 1 & $\frac{1}{2} \sin
\theta_p$ & $\approx 0$ & - \\ 
$\phi(1930) \rightarrow K \bar{K}^*$ & 4 & $\frac{1}{2\sqrt{2}}$ & $260\pm
109$ & * \\ 
$\phi(1930) \rightarrow \phi (1020)\eta$ & 1 & $\frac{1}{\sqrt{2}}\sin
\theta_p$ & $67 \pm 28$ & * \\ 
$\phi(1930) \rightarrow \phi (1020) \eta^{\prime }$ & 1 & $\frac{1}{\sqrt{2}}
\cos \theta_p$ & $\approx 0$ & * \\ \hline
\end{tabular}%
\end{table}
The obtained theoretical results are consistent with the experiments.
Especially, this is visible in the case of large decays. Yet, there are some
inconsistencies between our model and experimental data from PDG: the
theoretical decay width  $K^{\ast }(1410)\rightarrow K^{\ast }(892)$ is too
small in comparison to the experiment; on the contrary, $\rho
(1700)\rightarrow \bar{K}K$ is too large when compared to data. These
discrepancies are also visible when studying ratios between partial widths,
which are discussed in details together with other open issues in Ref. \cite%
{mpio}.\newline
Our results confirm the validity of the interpretation of both nonets as
predominantly quark$-$antiquark states. Note, however, that some
experimental results are quite old and new determination would be very
useful.

\subsection{Radiative decays of excited vector mesons}

\label{sec3.2}

We consider the radiative decays of excited vector mesons into a photon and
pseudoscalar meson $(\gamma P)$. We evaluate this type of process by using
Vector Meson Dominance (VMD) (see e.g. Ref. \cite{connell}). In practice, we
replace the vector field strength tensor according to the formula 
\begin{equation}
V_{\mu \nu }\rightarrow V_{\mu \nu }+\frac{e_{0}}{g_{\rho }}QF_{\mu \nu }%
\text{ ,}
\end{equation}%
where $F_{\mu \nu }$ is the field strength tensor of the photons. Moreover, $%
g_{\rho }=5.5\pm 0.5$ is the coupling constant which determines the
interaction, $e_{0}=\sqrt{4\pi \alpha }$ stands for the electric charge of
the proton, and $Q=diag\{\frac{2}{3},-\frac{1}{3},-\frac{1}{3}\}$ represents
the charge matrix of the quarks. In Tab. 4 we report the results for
radiative decays of both nonets of excited vector mesons. 
\begin{table}[h]
\caption{Amplitude's coefficients and decay widths of radiative decays of
radially $(V_{E})$ and~ orbitally~ $(V_{D})$ excited vector mesons.}%
\centering
\par
\renewcommand{\arraystretch}{1.25} 
\begin{tabular}{c|c|c|c|c|c}
\hline
\multicolumn{2}{c|}{} & \multicolumn{2}{c|}{$V_E$} & \multicolumn{2}{c}{$V_D$%
} \\ \hline
Decay channel & $\lambda$ & \multicolumn{4}{c}{Decay width [MeV]} \\ 
\cline{3-6}
&  & Theory & Exp. & Theory & Exp. \\ \hline
$\rho_{R} \rightarrow \gamma \pi $ & $\frac{1}{6}$ & $0.072 \pm 0.042$ & $-$
& $0.095 \pm 0.058$ & $-$ \\ 
$\rho_{R} \rightarrow \gamma \eta $ & $\frac{1}{2} \cos \theta_p$ & $0.23
\pm 0.14$ & $\sim0.2-1.5$ & $0.35 \pm 0.21$ & $-$ \\ 
$\rho_{R} \rightarrow \gamma \eta^{\prime }$ & $\frac{1}{2} \sin \theta_p$ & 
$0.056 \pm 0.033$ & $-$ & $0.13 \pm 0.08$ & $-$ \\ 
$K^*_{R} \rightarrow \gamma K $ & $\frac{1}{3}$ & $0.18 \pm 0.11$ & $< 0.0529
$ & $0.30 \pm 0.18$ & $-$ \\ 
$\omega_{R} \rightarrow \gamma \pi $ & $\frac{1}{2}$ & $0.60 \pm 0.36$ & $%
1.90 \pm 0.75$ & $0.78 \pm 0.47$ & $-$ \\ 
$\omega_{R} \rightarrow \gamma \eta $ & $\frac{1}{6} \cos \theta_p$ & $0.023
\pm 0.014$ & $-$ & $0.035 \pm 0.021$ & $-$ \\ 
$\omega_{R} \rightarrow \gamma \eta^{\prime }$ & $\frac{1}{6} \cos \theta_p$
& $0.0050 \pm 0.0030$ & $-$ & $0.012 \pm 0.007$ & $-$ \\ 
$\phi_{R} \rightarrow \gamma \eta $ & $\frac{1}{3} \sin \theta_p$ & $0.14
\pm 0.09$ & $+$ & $0.19 \pm 0.12$ & $*$ \\ 
$\phi_{R} \rightarrow \gamma \eta^{\prime }$ & $\frac{1}{3} \cos \theta_p$ & 
$0.076 \pm 0.045$ & $-$ & $0.13 \pm 0.08$ & $*$ \\ \hline
\end{tabular}%
\end{table}
We use the following notation: $\{\rho _{R},K_{R}^{\ast },\phi _{R},\omega
_{R}\}=\{\rho (1450),K^{\ast }(1410),\omega (1420),\phi (1680)\}$ stands for
the nonet of radially excited vector mesons and $\{\rho _{R},K_{R}^{\ast
},\phi _{R},\omega _{R}\}=\{\rho (1700),K^{\ast }(1680),\omega (1650),\phi
(1930)\}$ for orbitally excited vector mesons. We use the same notation as
in the case of strong decays to indicate which decay process was observed
and which one was not. Here, experimental data are rather poor, this is why
most of the results are predictions.

\subsection{Predictions for $\protect\phi(1930)$ resonance}

\label{sec3.3} Our approach allows us to make predictions for a not-yet
experimentally observed $s\bar{s}$ resonance: $\phi (???)=\phi (1930)$. The
existence of this missing state was discussed in the framework of the quark
model \cite{godfrey} and recently re-elaborated in Ref. \cite{wang}, based
on the similarity between non-strange and strange meson families. In order
to perform the calculations we have to know the mass of this putative state,
which we fix by making a simple estimate. We notice that the mass difference
between other member in both nonets of excited vector mesons is
approximately the same and reads $250$ MeV (this is due to the same dynamics
describing these resonances). We shall apply this relation also to our novel
state. By adding this difference to the mass of $\phi (1680)$, we get the
mass of the state $\phi (???)$: $1930$ MeV, hence $\phi (1930).$  By using
this assumption, we calculate the decay widths of this resonance, as it is
shown in Tab. 5. These results are predictions.

\begin{table}[h]
\caption{Predicted values of decay widths for the putative state $\protect%
\phi (1930)$. }\centering
\par
\renewcommand{\arraystretch}{1.25} 
\begin{tabular}{cc}
\hline
\multicolumn{2}{c}{Resonance $\phi (1930)$} \\ \hline
Decay channel & Decay width [Mev] \\ \hline
$\phi(1930) \rightarrow KK$ & $104 \pm 28$ \\ 
$\phi(1930) \rightarrow K\bar{K}^*$ & $260 \pm 109$ \\ 
$\phi(1930) \rightarrow \phi(1020)\eta$ & $67 \pm 28$ \\ 
$\phi(1930) \rightarrow \phi(1020) \eta^{\prime }$ & $\approx 0$ \\ 
$\phi(1930) \rightarrow \gamma \eta$ & $0.19 \pm 0.12$ \\ 
$\phi(1930) \rightarrow \gamma \eta^{\prime }$ & $0.13 \pm 0.08$ \\ \hline
\end{tabular}%
\end{table}
Our calculation shows that $\phi (1930)$ has a quite broad width ($\approx
400\pm 100$ MeV), which is probably the reason why it is still missing in
experiments. It is promising that in the near future two
photoproduction-based experiments, CLAS12 \cite{clas12} and GlueX \cite%
{gluex, gluex2, gluex3}, will take place at Jefferson Lab and will be able
to investigate that energy region.

\section{Conclusions}

\label{Sec.4} In this work we have studied strong and radiative decays of
two nonets of excited vector mesons: $\{\rho (1450),K^{\ast }(1410),\omega
(1420),\phi (1680)\}$, which predominantly corresponding to radially excited
vector mesons, and $\{\rho (1700),K^{\ast }(1680),\omega (1650),\phi (1930)\}
$, predominantly corresponding to orbitally excited vector mesons. In our
approach we used an effective QFT Lagrangian based on flavour symmetry. In
total, we evaluated 64 different decay processes which we compared to
experimental data from PDG (see results in tables 2-5). The overall
agreement of our theory with experimental results shows the validity of the
interpretation of these states as quark-antiquark objects. Moreover, we have
made predictions for an undiscovered $s\bar{s}$ resonance belonging to nonet
of orbitally excited vector mesons, which we named $\phi (1930)$. Our
calculations have shown that this missing state is broad. However, we expect
new data from CLAS12 and GlueX experiments, that could shed light on its
existence.

\section*{Acknowledgements}

The authors thank C. Reisinger for cooperation and acknowledge support from
the Polish National Science Centre (NCN) through the OPUS project no.
2015/17/B/ST2/01625.

\end{document}